\documentclass[twocolumn,showpacs,prc,superscriptaddress,amsmath,amssymb]{revtex4}

\usepackage{dcolumn}
\usepackage{bm}
\usepackage{url}
\usepackage{color}
\usepackage[T1]{fontenc}
 \usepackage{epsfig}
\usepackage{graphicx}
\usepackage{calc}
\usepackage{picture}

\begin{document}

\preprint{APS/123-QED}

\title{Properties of $^{187}$Ta revealed through isomeric decay}

\author{P.M.~Walker}\email{p.walker@surrey.ac.uk}\affiliation{Department of Physics, University of Surrey, Guildford, GU2 7XH, UK}
\author{Y.~Hirayama}\affiliation{Wako Nuclear Science Center (WNSC), Institute of Particle and Nuclear Studies (IPNS), High Energy Accelerator Research Organization (KEK), Wako, Saitama 351-0198, Japan}
\author{G.J.~Lane}\affiliation{Department of Nuclear Physics, RSPhys, Australian National University, Canberra ACT 2601, Australia}
\author{H.~Watanabe}\affiliation{School of Physics, and International Research Center for Nuclei and Particles in Cosmos, Beihang University, Beijing 100191, China}\affiliation{Nishina Center for Accelerator-Based Science, RIKEN, Wako, Saitama 351-0198, Japan}
\author{G.D.~Dracoulis}\affiliation{Department of Nuclear Physics, RSPhys, Australian National University, Canberra ACT 2601, Australia}
\author{M.~Ahmed}\affiliation{Wako Nuclear Science Center (WNSC), Institute of Particle and Nuclear Studies (IPNS), High Energy Accelerator Research Organization (KEK), Wako, Saitama 351-0198, Japan}\affiliation{University of Tsukuba, Tsukuba, Ibaraki 305-0006, Japan}
\author{M.~Brunet}\affiliation{Department of Physics, University of Surrey, Guildford, GU2 7XH, UK}
\author{T.~Hashimoto}\affiliation{Rare Isotope Science Project, Institute for Basic Science (IBS), Daejeon 305-811, Republic of Korea}
\author{S.~Ishizawa}\affiliation{Nishina Center for Accelerator-Based Science, RIKEN, Wako, Saitama 351-0198, Japan}\affiliation{Graduate School of Science and Engineering, Yamagata University, Yamagata 992-8510, Japan}\affiliation{Wako Nuclear Science Center (WNSC), Institute of Particle and Nuclear Studies (IPNS), High Energy Accelerator Research Organization (KEK), Wako, Saitama 351-0198, Japan}
\author{F.G.~Kondev}\affiliation{Physics Division, Argonne National Laboratory, Lemont, Illinois 60439, USA}
\author{Yu.A.~Litvinov}\affiliation{GSI Helmholtzzentrum f\"ur Schwerionenforschung, 64291 Darmstadt, Germany}
\author{H.~Miyatake}\affiliation{Wako Nuclear Science Center (WNSC), Institute of Particle and Nuclear Studies (IPNS), High Energy Accelerator Research Organization (KEK), Wako, Saitama 351-0198, Japan}
\author{J.Y.~Moon}\affiliation{Rare Isotope Science Project, Institute for Basic Science (IBS), Daejeon 305-811, Republic of Korea}
\author{M.~Mukai}\affiliation{University of Tsukuba, Tsukuba, Ibaraki 305-0006, Japan}\affiliation{Wako Nuclear Science Center (WNSC), Institute of Particle and Nuclear Studies (IPNS), High Energy Accelerator Research Organization (KEK), Wako, Saitama 351-0198, Japan}\affiliation{Nishina Center for Accelerator-Based Science, RIKEN, Wako, Saitama 351-0198, Japan}
\author{T.~Niwase}\affiliation{Wako Nuclear Science Center (WNSC), Institute of Particle and Nuclear Studies (IPNS), High Energy Accelerator Research Organization (KEK), Wako, Saitama 351-0198, Japan}\affiliation{Nishina Center for Accelerator-Based Science, RIKEN, Wako, Saitama 351-0198, Japan}\affiliation{Department of Physics, Kyushu University, Nishi-ku, Fukuoka 819-0395, Japan}
\author{J.H.~Park}\affiliation{Rare Isotope Science Project, Institute for Basic Science (IBS), Daejeon 305-811, Republic of Korea}
\author{Zs.~Podoly\'ak}\affiliation{Department of Physics, University of Surrey, Guildford, GU2 7XH, UK}
\author{M.~Rosenbusch}\affiliation{Wako Nuclear Science Center (WNSC), Institute of Particle and Nuclear Studies (IPNS), High Energy Accelerator Research Organization (KEK), Wako, Saitama 351-0198, Japan}
\author{P.~Schury}\affiliation{Wako Nuclear Science Center (WNSC), Institute of Particle and Nuclear Studies (IPNS), High Energy Accelerator Research Organization (KEK), Wako, Saitama 351-0198, Japan}
\author{M.~Wada}\affiliation{Wako Nuclear Science Center (WNSC), Institute of Particle and Nuclear Studies (IPNS), High Energy Accelerator Research Organization (KEK), Wako, Saitama 351-0198, Japan}\affiliation{University of Tsukuba, Tsukuba, Ibaraki 305-0006, Japan}
\author{X.Y.~Watanabe}\affiliation{Wako Nuclear Science Center (WNSC), Institute of Particle and Nuclear Studies (IPNS), High Energy Accelerator Research Organization (KEK), Wako, Saitama 351-0198, Japan}
\author{W.Y.~Liang}\affiliation{School of Physics and State Key Laboratory of Nuclear Physics and Technology, Peking University, Beijing 100871, China}
\author{F.R.~Xu}\affiliation{School of Physics and State Key Laboratory of Nuclear Physics and Technology, Peking University, Beijing 100871, China}

\begin{abstract}
Mass-separated $^{187}$Ta$_{114}$ in a high-spin isomeric state has been produced for the first time by multi-nucleon transfer reactions, employing an argon gas stopping cell and laser ionisation. Internal $\gamma$~rays revealed a $T_{1/2} = 7.3$$\pm$0.9~s isomer at 1778$\pm$1~keV, which decays through a rotational band with perturbations associated with the approach to a prolate-oblate shape transition. Model calculations show less influence from triaxiality compared to  heavier elements in the same mass region. The isomer decay reduced $E2$ hindrance factor, $f_\nu = 27$$\pm$1, supports the interpretation that axial symmetry is approximately conserved.
\end{abstract}
  

\date{\today}

\maketitle


Progress in the study of heavy, neutron-rich nuclei is leading the way towards a deeper understanding of cosmic elemental abundances. For example, 
the $A \approx 195$ solar abundance peak is determined by the properties of unstable nuclei along the 
$r$-process (rapid neutron capture) path of stellar nucleosynthesis, which follows the $N=126$ closed neutron shell up to $Z \approx 73$ \cite{Ka17}. Pinning down the structure of the $N \approx 126$ nuclei, and ultimately reaching the $r$-process path itself, are basic objectives of several new accelerator facilities, both planned and under construction \cite{USLRP,EULRP,Ka19,Ho19}. Nevertheless, difficulties center on the low production rates and the lack of selective reaction mechanisms, so that additional experimental advances are needed \cite{Ka19,Ho19}. A possibility for half-lives longer than about 100~ms is `isotope separation', where the nuclear reaction products are stopped, vaporised, ionised, mass separated, and transported to a low-background measurement station. However, the refractory chemistry of key elements, from hafnium ($Z=72$) to iridium ($Z=77$), makes them hard to vaporise \cite{Ko07}. An essential requirement is therefore to develop a suitable gas stopping arrangement for the reaction products \cite{Hi15,Sa20}.

A promising additional feature of isotope separation is that it may be used with long-lived nuclear excited states (isomers) \cite{Wa99,Dr16,Wa20} whose $\gamma$-ray decay can enable the nuclear structure to be studied at high angular momentum, giving sensitivity to rotational and shape degrees of freedom. While the occurrence of sufficiently long-lived, high-spin isomers is rare \cite{Ja15}, discoveries \cite{Re10} in the Experimental Storage Ring (ESR) at GSI in Germany, have opened a special opportunity in the neutron-rich hafnium and tantalum ($Z=73$) isotopes.

We now report the first successful production and separation of low-energy beams of neutron-rich tantalum isotopes and isomers. The production process exploits multi-nucleon transfer (MNT) reactions which have been shown to be effective for making neutron-rich nuclei \cite{Wat15,Li19,De19}. In the present work, the large angular momentum transfer in MNT reactions is a vital aspect for the formation of high-spin isomers.
We focus on the internal decay of an isomer in $^{187}$Ta, which reveals a perturbed rotational band and sheds light on the nuclear structure associated with a prolate-oblate shape transition.

The existence of $^{187}$Ta was first established at GSI from high-energy projectile fragmentation of $^{197}$Au \cite{Be99}. Subsequently, it was observed with high mass resolution in the ESR using the same reaction. In the form of `bare' ions with all atomic electrons removed, the $^{187}$Ta ground state ($gs$) and two isomeric states ($m1$ and $m2$) were identified and their masses and half-lives measured, yielding excitation energies of 1789$\pm$13 keV and 2935$\pm$14~keV, respectively \cite{Re10}. The measured half-lives were 2.3$\pm$0.6~min, 22$\pm$9~s and $>$5 min for the $gs$, $m1$ and $m2$ states, respectively. Furthermore, without observing the decay radiations, it could be determined that there were both $\beta$ and $\gamma$ decays from the $m1$ isomer, but no other spectroscopic information was obtained. The key finding now reported  is the detailed $\gamma$-ray decay pathway from the $m1$ isomer to the $gs$ of $^{187}$Ta. The first observation of $\gamma$-rays following the $\beta$ decay of the  $gs$  will be reported elsewhere,
as will the tentative identification of decays associated with the $m2$ isomer. 

The experiment was performed at the RIKEN Nishina Center in Saitama, Japan, with the recently commissioned KEK Isotope Separation System (KISS) facility \cite{Hi20,Wat20}. This is the first facility of its kind, capable of stopping heavy-ion reaction products in a high-pressure (80 kPa) argon gas cell, performing laser resonant ionisation for element ($Z$) selectivity, and achieving mass ($A$) separation of the electrostatically extracted, singly charged, 20 keV ions in a dipole magnet with a resolving power $A/\Delta A = 900$. In the present work, the $^{187}$Ta ions were produced by  MNT reactions of a 50 particle-nA beam of $^{136}$Xe at 7.2 MeV per nucleon, delivered by the RIKEN Ring Cyclotron. The beam was incident on a 5~$\mu$m thick natural tungsten target (28\% $^{186}$W) at the entrance to the argon gas cell. 

The 20 keV secondary beam of laser-ionised tantalum \cite{Hi19} was mass separated  ($\approx$1.5 ions/s of $^{187}$Ta) and transported to a moving-tape collection point, surrounded by a low-background, 32-element 
gas proportional counter with 80\% of 4$\pi$ solid angle for $\beta$ particles and conversion electrons \cite{Mu20}, and four Super Clover germanium $\gamma$-ray detectors with a total absolute full-energy-peak efficiency of 15\% at 150~keV. In the analysis, $e$$-$$\gamma$ coincidences were always required, as they were very effective at removing background $\gamma$-ray events. 
Furthermore, the 32 elements of the gas counter were arranged in two concentric layers, and, by requiring events only in the first layer, isomeric cascades could be preferentially selected \cite{Wat20,Mu20}.
The tape transport was operated with equal beam-on/beam-off periods, with the radioactivity  moved to a shielded location at the end of each cycle. The chosen beam-on periods were 30 s, 300 s and 1800 s, with five days of data taking. During the beam-off periods at the $e$$-$$\gamma$ spectroscopy station, the beam was electrostatically deflected to a multi-reflection time-of-flight (MR-TOF) device for high-resolution mass analysis \cite{Sc20}.

\begin{figure}[tb]
\vspace{-7.5cm}
\hspace{-0.1cm}
\epsfig{file=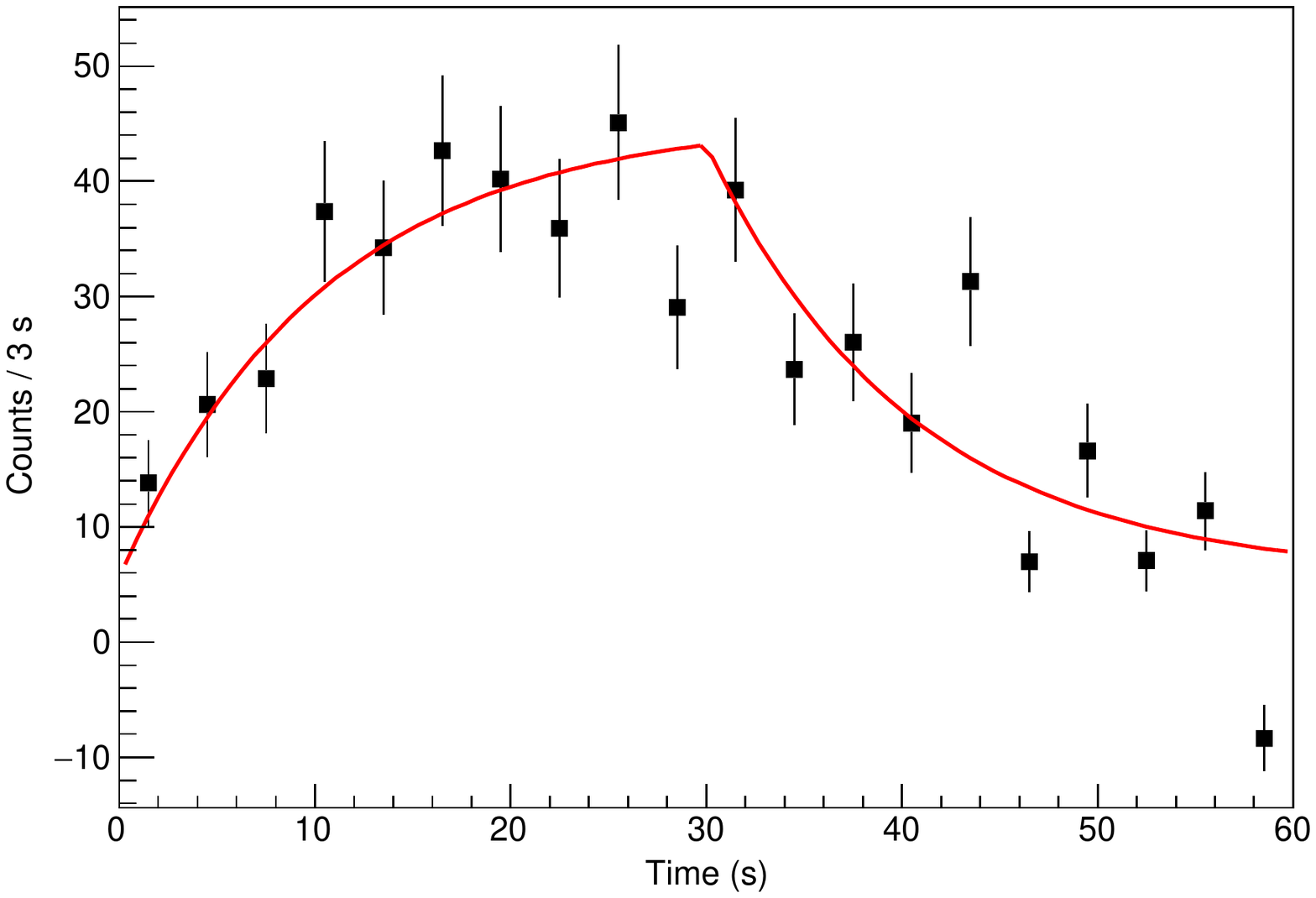,scale=0.45}
\begin{center}
\vspace{-0.5cm}
\begin{minipage}[t]{8.5 cm}
\caption{Sum of $\gamma$-ray time spectra for transitions involved in the decay of $^{187}$Ta$^{m1}$. The line through the data is a log-likelihood fit to the growth (beam on) and decay (beam off) periods, each of 30 s duration, yielding a half-life of 7.3$\pm$0.9~s. \label{time}}
\end{minipage}
\end{center}
\end{figure}

First the $m1$ isomer half-life is discussed. The half-life derived from the ESR data for bare ions represents an upper half-life limit for neutral atoms, because conversion coefficients can be very large (greater than 100) for low-energy $\gamma$ rays \cite{Ki08}, yet electron conversion cannot take place in bare ions. Therefore, the present measurement searched for previously unassigned $\gamma$-ray transitions with $T_{1/2} \leq 22$~s. As illustrated in Fig.~\ref{time}, transitions were identified that yielded a half-life of 7.3$\pm$0.9 s. The time spectrum comes from the addition of the time evolution of ten different $\gamma$-ray transition intensities. For individual transitions, the half-life uncertainties are large, but by simple inspection of the time-gated $\gamma$-ray spectra it is clear that their half-lives are all less than 20 s, and no other $^{187}$Ta half-lives of less than one minute are known.

\begin{figure}[tb]
\epsfig{file=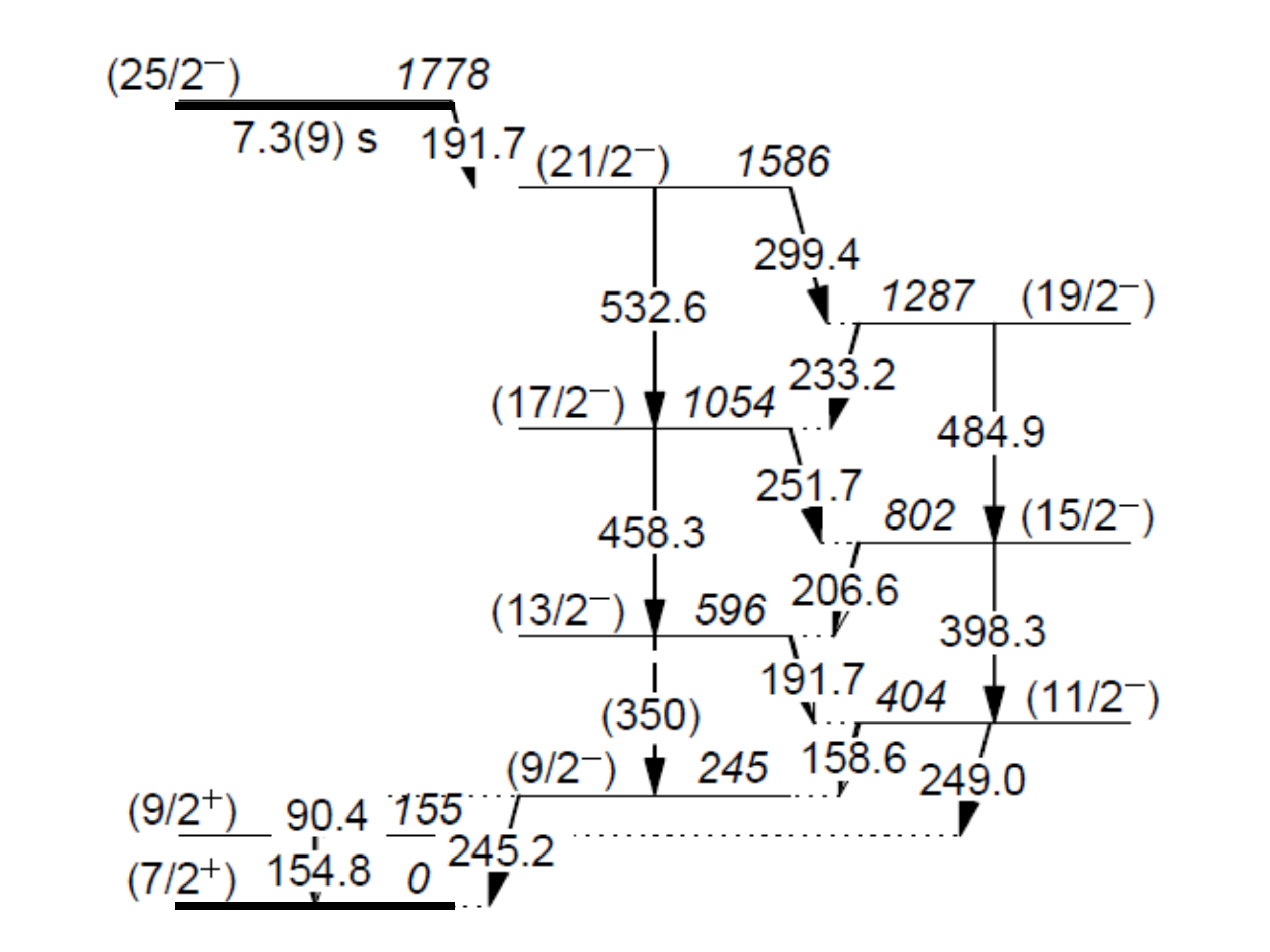,scale=0.25}
\begin{center}
\vspace{-0.8cm}
\begin{minipage}[t]{8.5 cm}
\caption{Level scheme for $^{187}$Ta obtained from the decay of its $T_{1/2} = 7.3$$\pm$0.9 s isomeric state at 1778 keV. \label{ta187-ls}}
\end{minipage}
\end{center}
\end{figure}

\begin{figure}[tb]
\vspace{-11.5cm}
\hspace{-0.1cm}
\epsfig{file=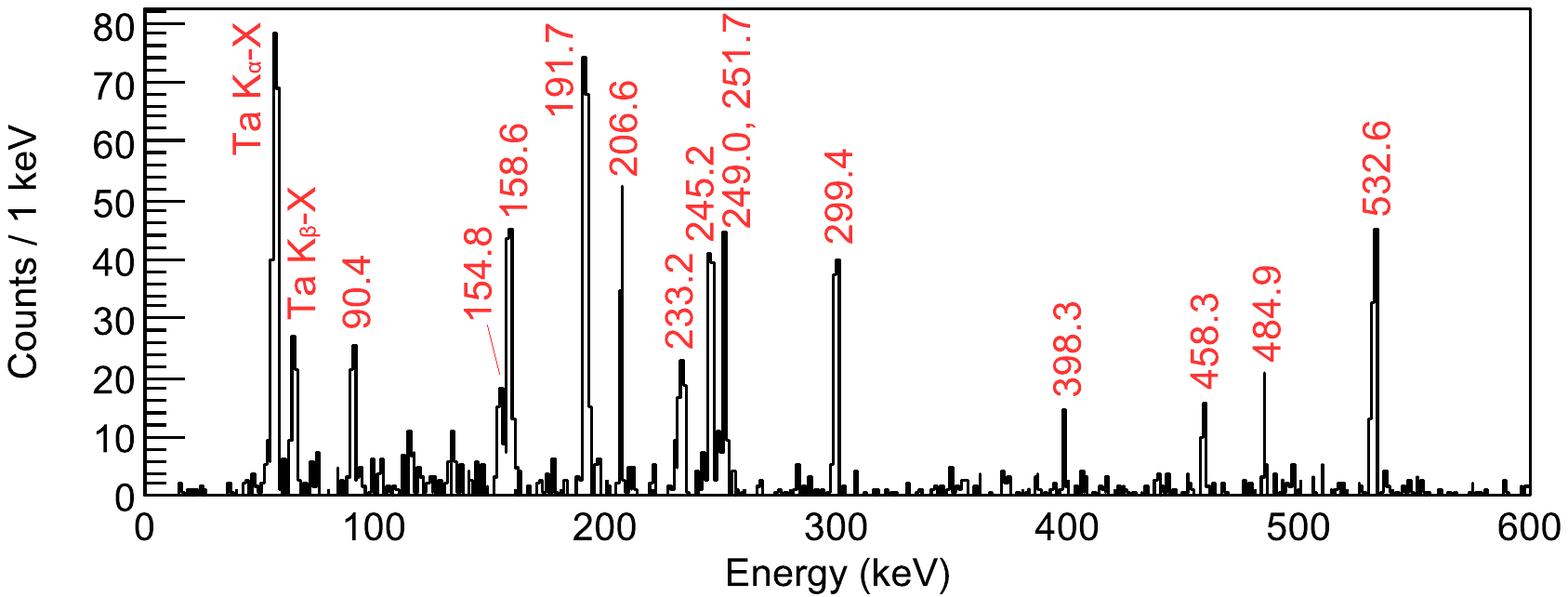,scale=0.5}
\begin{center}
\vspace{-0.5cm}
\begin{minipage}[t]{8.5 cm}
\caption{Sum of $\gamma$-$\gamma$-coincidence energy spectra, illustrating transitions involved in the decay of  $^{187}$Ta$^{m1}$. \label{gamma}}
\end{minipage}
\end{center}
\end{figure}

Having identified a number of $\gamma$-ray transitions associated with the relatively short half-life of 7.3~s,  the $\gamma$-ray coincidence relationships ($\pm$300 ns) enabled them to be unambiguously ordered in a cascade, as shown in Fig.~\ref{ta187-ls}. The $\gamma$-ray sum-coincidence spectrum is illustrated in Fig.~\ref{gamma}, where the tantalum x rays show that the isomer is undergoing internal decay. 
The level structure is independently confirmed by Gammasphere data from several years ago \cite{La09}, also using MNT reactions, taken at the Argonne National Laboratory, USA.  Previously unplaced $^{187}$Ta transitions and their coincidence relationships have now been observed during the beam-off periods of the Gammasphere beam pulsing. 

From the Gammasphere data, the $\gamma$-ray intensities, combined with theoretical conversion coefficients \cite{Ki08}, enable the transition multipole characters to be deduced, assuming rotational properties for the band based on the 245~keV level.  We discuss in more detail only the 191.7~keV doublet. It proved not to be possible to distinguish between the energies of these two transitions (with uncertainties of $\pm$0.5~keV) but their intensities can be separately determined. Thus the conversion coefficient of the isomeric 191.7~keV transition is found to be $\alpha _T = 0.57$$\pm$0.24, which indicates $M1$ or $E2$ character. The theoretical conversion coefficients \cite{Ki08} are $\alpha _T = 0.07$, 0.69, 0.34, 3.72 and 2.69 for $E1$, $M1$, $E2$, $M2$ and $E3$, respectively. However, an $M1$ assignment would indicate the possibility of a competing $E2$ transition to the 1287~keV band member, which is unobserved. Therefore, a tentative $E2$ assignment is given for the isomeric 191.7~keV transition, and hence $I^\pi = (25/2^-)$  for the isomer itself.

%
%

Comparison with the ESR data \cite{Re10} is instructive. Those data independently establish the energy difference between the isomer and the $gs$ to be 1789$\pm$13~keV, in good agreement with the present more-accurate value of 1778$\pm$1~keV.
The ESR measurements \cite{Re10,Re12} also identified  five $\beta$-decay events (as well as eight $\gamma$-decay events) from $^{187}$Ta$^{m1}$ but, in the present work, no $\beta$-decay branch could be confirmed. Considering both data sets, and that electron conversion is not possible for bare ions in the ESR, we estimate conservatively that the neutral atom has a $\beta$-decay branching ratio $<$40\%.

The structure of the $gs$ of $^{187}$Ta is indicated by the systematic observation \cite{Ja90} that the $gs$  of the lighter tantalum isotopes ($175 \leq A \leq 185$) is formed by the spin $I^\pi = \Omega ^\pi = 7/2^+$, 7/2$^+$[404] Nilsson configuration, where $\Omega$ is the spin projection on the nuclear symmetry axis. The same configuration assignment for $^{187}$Ta is supported in the present work 
by the observed $\beta$ decay of the $gs$ to states in $^{187}$W. However, parentheses are used for all spin and parity assignments in Fig.~\ref{ta187-ls} on account of the provisional $gs$ assignment.
The next-lowest intrinsic state is expected \cite{Ja90} to be the 9/2$^-$[514] configuration. This is entirely consistent with the observed structure, which is itself very similar to the corresponding structure in $^{185}$Ta \cite{La09}. Once the $gs$ spin and parity are taken to be 7/2$^+$, as for the 7/2$^+$[404] configuration, all the other spins and parities in Fig.~\ref{ta187-ls} are determined by the $\gamma$-ray intensities and implied conversion coefficients, as discussed above.

%


The isomer configuration comes from multi-quasiparticle calculations. Reed et al.~\cite{Re10,Re12}, assigned a $K^\pi = 27/2^+$  one-proton/two-neutron configuration $\{ \pi 9/2^- [514]\otimes \nu 11/2^+ [615]\otimes \nu 7/2^- [503]\}$, where $K$ is the sum of the individual $\Omega$ values. This would be compatible with the deduced level scheme if the 191.7~keV transition from the isomer were of $E3$ character, but that is ruled out by the deduced conversion coefficient.
We have performed two new sets of calculations with a Woods-Saxon potential and universal parameters \cite{Cw87}, firstly  multi-configuration calculations \cite{Ja95,Ko97,Ha20} with fixed deformation \cite{Mo16}, including residual interactions, and secondly configuration-constrained,  potential-energy-surface (PES) calculations \cite{Xu98}. The calculations suggest two competing $K^\pi = 25/2^-$ configurations: $\{ \pi 7/2^+ [404]\otimes \nu 11/2^+ [615]\otimes \nu 7/2^-[503]\}$, and $\{ \pi 9/2^- [514]\otimes \nu 9/2^- [505]\otimes \nu 7/2^- [503]\}$, but we are unable to distinguish between them experimentally. The PES calculations give deformation parameters ($\beta _2,\gamma,\beta _4$) = (0.208,\,0$^\circ$,$-$0.076) and (0.189,\,0$^\circ$,$-$0.062) respectively. With regard to isomer decay rates (discussed later) the $\gamma = 0^\circ$ axial symmetry is a significant feature.

\begin{figure}[tb]
\vspace{-2.0cm}
\hspace{-0.3cm}
\epsfig{file=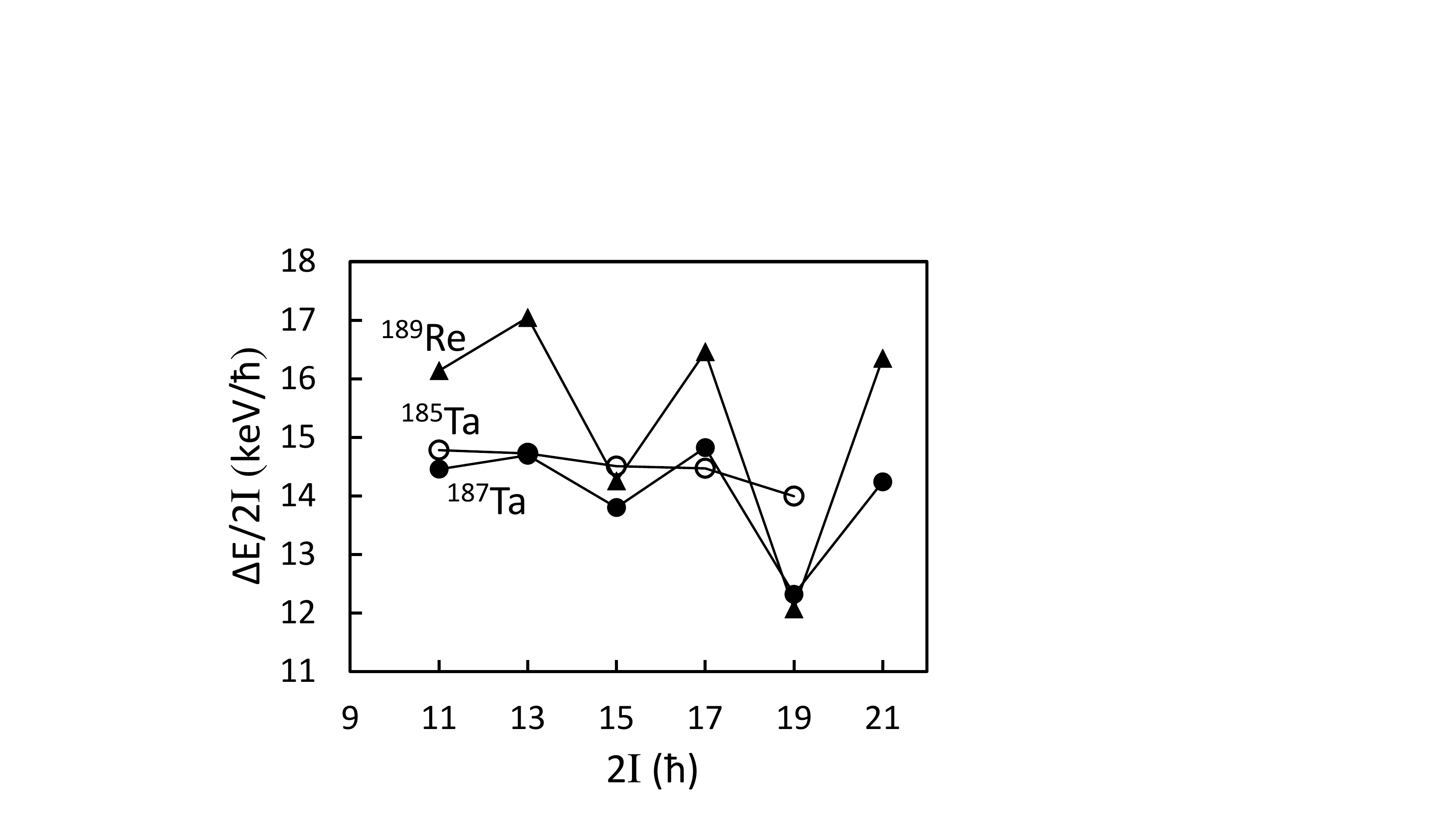,scale=0.35}
\begin{center}
\vspace{-0.8cm}
\begin{minipage}[t]{8.5 cm}
\caption{Signature splitting as a function of angular momentum for 9/2$^-$[514] bands in $^{185}$Ta \cite{La09}, $^{187}$Ta (this work) and $^{189}$Re \cite{Re16}. The energy, $\Delta E$, is for $I \rightarrow I-1$ transitions. Circles represent tantalum isotopes ($Z=73$) and filled symbols represent $N=114$ isotones. \label{staggering}}
\end{minipage}
\end{center}
\end{figure}

Considering the rotational states populated by the isomer decay, it is instructive to analyse the $I\rightarrow I-1$ transition energies. A rotational band with a constant moment of inertia, $J$, follows the well-known formula for the rotational energy, $E_R = \frac{\hbar ^2}{2J}[I(I+1)-K^2]$, so that the energy change, $\Delta E(I\rightarrow I-1)$, obeys the expression $\Delta E/2I = \frac{\hbar ^2}{2J}$. In order to assess the regularity of a rotational band, it is common practice to plot this as a function of $I$, as shown in Fig.~\ref{staggering} for the 9/2$^-$[514] bands in $^{187}$Ta, its isotope $^{185}$Ta \cite{La09},  and its isotone $^{189}$Re \cite{Re16}. For  $^{185}$Ta, the behavior is monotonic and slightly decreasing. This indicates a gently increasing moment of inertia, which is the normal behavior. By contrast, there are significant oscillations, or `staggering', for $^{187}$Ta, which become substantially larger in $^{189}$Re. This staggering effect has recently been studied in the 9/2$^-$[514] bands of $^{187,189,191}$Re, both experimentally and with triaxial particle-plus-rotor model calculations \cite{Re16}. It was concluded that the triaxiality parameter (which ranges from $\gamma = 0^\circ$ for prolate shape to  60$^\circ$ for oblate shape) takes values of 5$^\circ$, 18$^\circ$ and 25$^\circ$, respectively, in the three rhenium isotopes. Since the staggering in the $^{187}$Ta band is intermediate between that in $^{189}$Re and $^{187}$Re \cite{Re16}, it can be estimated that its triaxiality value is $\gamma \approx 10^\circ$. It can also be observed in Fig.~\ref{staggering} that the $\Delta E/2I$ value is, on average, lower for $^{187}$Ta than for $^{189}$Re, indicating a higher moment of inertia for $^{187}$Ta, and hence a larger $\beta_2$ deformation.

\begin{figure}[tb]
\vspace{-8.5cm}
\epsfig{file=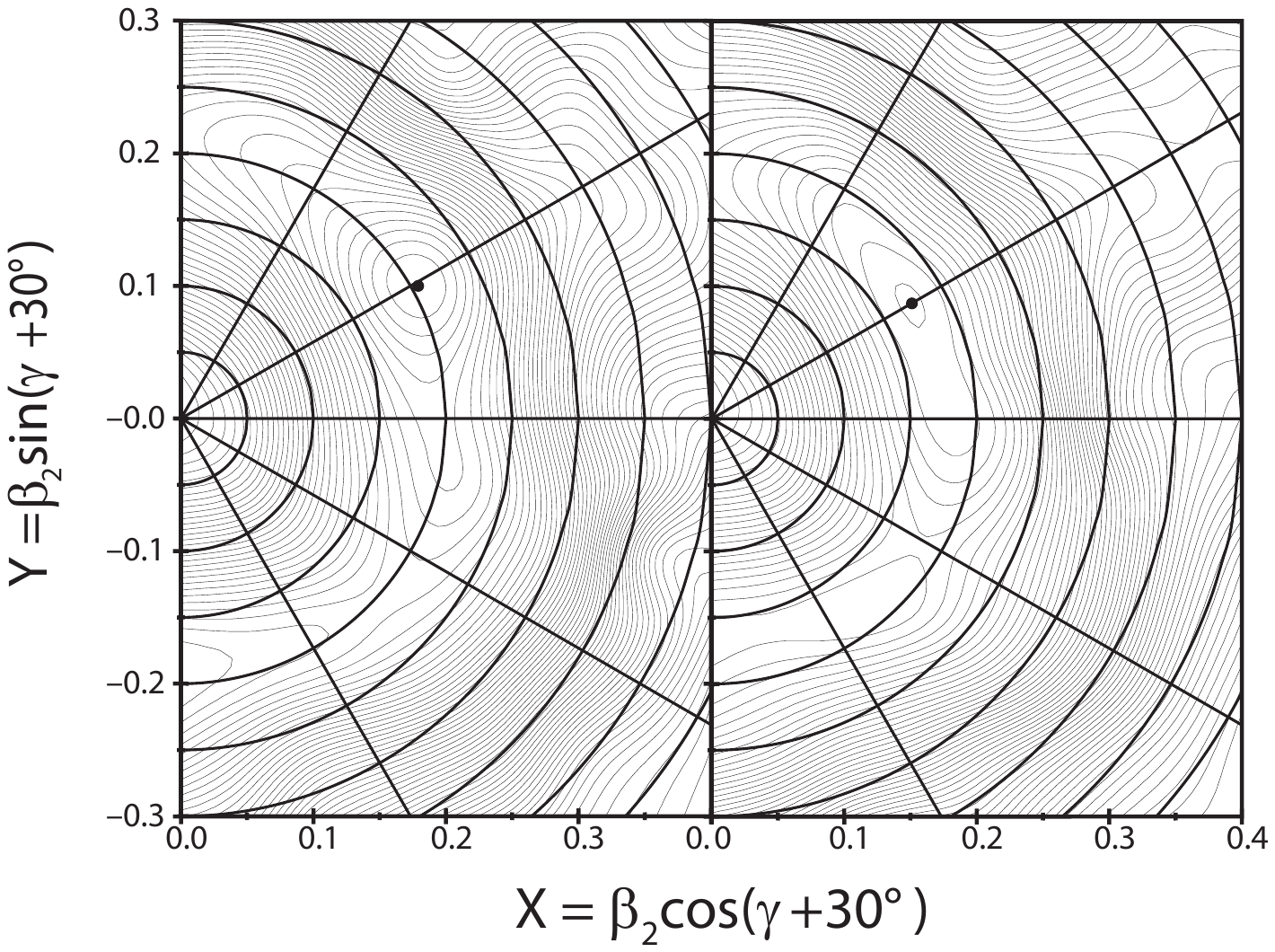,scale=0.47}
\begin{center}
\vspace{-0.5cm}
\begin{minipage}[t]{8.5 cm}
\caption{TRS calculations for the 9/2$^-$[514] bands in $^{187}$Ta (left) and $^{189}$Re (right) at $\hbar \omega = 0.15$~MeV ($I\approx 5$$\hbar$). The energy minima (dots) are at 
($\beta _2,\gamma,\beta _4$) = (0.205,\,0$^\circ$,$-$0.079) for $^{187}$Ta and (0.174,\,0$^\circ$,$-$0.060) for $^{189}$Re.
%
%
Energy contours are at 200~keV intervals. \label{TRS}}
\end{minipage}
\end{center}
\end{figure}

In the present work, Total Routhian Surface (TRS) calculations \cite{Sa94,Sa95,Xu00} have been carried out for the 9/2$^-$[514] bands of $^{187}$Ta and $^{189}$Re. As illustrated in Fig.~\ref{TRS} these show axially symmetric, prolate minima. The minimum is shallower for $^{189}$Re and the contour lines extend in the axially asymmetric direction, particularly in the $\gamma < 0^\circ$ direction. This is consistent with a greater role of axial asymmetry in $^{189}$Re, relative to $^{187}$Ta. Compared to the particle-plus-rotor calculations \cite{Re16}, the present work suggests the significant role of dynamic, rather than static, $\gamma$ deformation at low spin. However, at high spin ($I \approx 18$) the TRS calculations show a dramatic change to oblate rotation, similar to that predicted in, for example, the hafnium isotopes \cite{Xu00}.

An important aspect here is that $N \approx 116$ is the critical point for a $gs$ prolate-oblate shape transition with increasing neutron number. This has been well studied in the higher-$Z$ elements, especially osmium ($Z=76$) and platinum ($Z=78$), where the prolate-oblate transition is achieved by passing through triaxial shapes \cite{Dr16,Cl86,Wu96,Jo03,Ro09,No11}. However, calculations such as those of Robledo et al.~\cite{Ro09} and the present work predict that, as $Z$ decreases, the $\beta_2$ deformation increases and triaxiality  plays a reduced role. 
%
%
We have here presented experimental evidence, through the reduced staggering behavior in the 9/2$^-$[514] band of $^{187}$Ta, compared to that in its $N=114$ isotone $^{189}$Re, that this is indeed the case.

Further information about the triaxiality comes from the half-life of $^{187}$Ta$^{m1}$ ($T_{1/2}=7.3$$\pm$0.9 s) which is long because the decay transition is `$K$ forbidden', with $\Delta K > \lambda$, where $\Delta K$ is the change in $K$ value, and $\lambda$ is the angular momentum carried by the transition. The degree of forbiddenness is defined as $\nu = \Delta K - \lambda$, and the reduced hindrance is $f_\nu =( T_{1/2}^\gamma/T_{1/2}^W)^{1/\nu}$, where $T_{1/2}^\gamma$ is the partial $\gamma$-ray half-life and $T_{1/2}^W$ is the corresponding Weisskopf estimate \cite{Dr16}. In the present case of $E2$ decay from $^{187}$Ta$^{m1}$, we find that $f_\nu = 27$$\pm$1, which is a substantial value \cite{Dr16,Ko15}. However, compared to the $E2$ decay of $^{185}$Ta$^{m1}$, with $f_\nu = 71$ \cite{La09}, the $f_\nu = 27$ value indicates significant $K$ mixing, qualitatively consistent with the greater staggering in the populated rotational band. Comparison with the equivalent $E2$ isomeric decay in $^{189}$Re is not straightforward, because the branching intensities are not specified \cite{Re16}. Nevertheless, a much smaller $f_\nu$ value is evident, indicating a substantial loss of axial symmetry.
%
%
Therefore, this analysis supports the calculations (\cite{Ro09} and the present work) that show better axial symmetry for lower-$Z$ elements in the $N\approx 116$ prolate-oblate shape transition region.

In summary, the excited-state structure of $^{187}$Ta$_{114}$ has been revealed for the first time through its $m1$ isomer decay, with $T_{1/2}=7.3$$\pm$0.9 s. Despite the close approach to  $N\approx 116$, which is the critical point for the predicted $gs$ prolate-oblate shape transition, the reduced hindrance for the $E2$ isomeric decay remains substantial, with $f_\nu = 27$$\pm1$, indicating that $K$ is approximately conserved, and therefore that axial symmetry is not strongly violated. Nevertheless, weak violation of axial symmetry is indicated by the observed staggering in the 9/2$^-$[514] rotational band that is populated through the isomer decay. Comparison with the rhenium isotone, $^{189}$Re, supports  calculations showing that axial symmetry is better conserved, through the  $N\approx 116$ shape transition region, for the lower-$Z$ nuclei. 
The new capability to produce low-energy beams of neutron-rich tantalum isotopes and isomers demonstrates the power of the gas-stopping technique for nuclear structure studies of exotic neutron-rich nuclei, even with refractory elements. This marks a milestone on the way to the exploration of nuclei predicted to have well-deformed oblate ground states, and ultimately to $N \approx 126$ $r$-process nuclei.


{\bf Acknowledgements:}
This experiment was performed at the RI Beam Factory
operated by RIKEN Nishina Center and CNS, University of
Tokyo. The authors thank the RIKEN accelerator staff
for their support. This work was funded in part by 
grant Nos.
JP23244060, JP24740180, JP26247044, JP15H02096, 
JP17H01132,  JP17H06090, and JP18H03711 from JSPS KAKENHI;
ST/L005743/1 from U.K.~STFC; 11921006 and 11835001 from NSFC; 682841 ``ASTRUm'' from ERC (Horizon 2020); and DE-AC02-06CH11357  from U.S.~Department of Energy (Office of Nuclear Physics).


\end{document}